%% file: Korneeva_PRB_submission.tex
\documentclass[amsmath,amssymb,aps,prb,10pt,twocolumn,superscriptaddress]{revtex4}
\usepackage{graphicx}
\usepackage{dcolumn}
\usepackage{bm}
\usepackage{xcolor}
\usepackage{soul,color}
\usepackage{textgreek}
\usepackage[breaklinks=true,colorlinks=true,urlcolor=blue,citecolor=blue]{hyperref}

\begin{document}

\title{Different single photon response of wide and narrow superconducting MoSi strips}

\author{Yu.~P.~Korneeva}
\affiliation{Physics Department, Moscow Pedagogical State University, Moscow, 119991 Russia}
\email{korneeva@rplab.ru}
\author{N.~N.~Manova}
\author{I.~N.~Florya}
\affiliation{Physics Department, Moscow Pedagogical State University, Moscow, 119991 Russia}
\author{M.~Yu.~Mikhailov}
\affiliation{B. Verkin Institute for Low Temperature Physics and Engineering of the National Academy of Sciences of Ukraine,  Kharkiv 61103, Ukraine}
\author{O.~V.~Dobrovolskiy }
\affiliation{Faculty of Physics, University of Vienna, 1090 Vienna, Austria}
\affiliation{Physics Department, V. Karazin Kharkiv National University, Kharkiv  61022, Ukraine}
\author{A.~A.~Korneev}
\affiliation{Physics Department, Moscow Pedagogical State University, Moscow, 119991 Russia}
\affiliation{National Research University Higher School of Economics, Moscow, 101000 Russia}
\author{D.~Yu.~Vodolazov}
\affiliation{Physics Department, Moscow Pedagogical State University, Moscow, 119991 Russia}
\affiliation{Institute for Physics of Microstructures, Russian Academy of Sciences, Nizhny Novgorod, GSP-105, Russia}


\begin{abstract}
The photon count rate of superconducting single photon detectors made of MoSi films shaped as a $2\,\mu$m-wide strip and a $115$\,nm-wide meander stripline is studied experimentally as a function of the dc biasing current at different values of the perpendicular magnetic field. For the wide strip a crossover current $I_\textrm{cross}$ is observed, below which the photon count rate increases with increasing magnetic field and above which it decreases. This behavior contrasts with the narrow MoSi meander for which no crossover current is observed, thus suggesting different photon detection mechanisms in the wide and narrow strips. Namely, we argue that in the wide strip the absorbed photon destroys superconductivity \emph{locally} via the vortex-antivortex mechanism for the emergence of resistance while in the narrow meander superconductivity is destroyed across the whole stripline, forming a hot belt. Accordingly, the different photon detection mechanisms associated with vortices and the hot belt stipulate the qualitative difference in the dependence of the photon count rate on the magnetic field.
\end{abstract}

\maketitle

\section{Introduction}
Recently, it was reported that single photon response in NbN thin films occurring in the well-known narrow and long meandering nanowires \cite{Gol01apl,Nat12sst}, also occurs in a few micron wide and tapered constrictions \cite{Kor18pra}. This experimental finding is an indirect indication of the applicability of the vortex-antivortex assisted mechanism for single-photon counting, which was introduced in the theoretical work by Zotova and Vodolazov (ZV) \cite{Zot12prb,Vod17pra}. In the ZV model, the only requirement is that the superconducting bridge carries over its width a uniform bias current whose value is close to the critical pair-breaking one. Thus, this requirement ensures that the bridge operates in a ``local'' regime. Namely, whenever a photon hits the bridge, creating a local non-equilibrium state (which is often called ``hot spot''), the supercurrent density distribution is changed inside and close to the ``impact point'' and it is not perturbed further away. Then, a vortex-antivortex pair is created and driven by the Lorentz force causing the emergence of dissipation in the superconductor. In this mechanism, each point of the bridge contributes to photon counting and the intrinsic detection efficiency is predicted to reach almost $100 \%$. Specifically, the ZV model requires the bias current in the bridge to roughly exceed a half of the depairing current. Previously, these conditions were experimentally satisfied in wide and short NbN bridges studied by Korneeva \emph{et al} \cite{Kor18pra}. The experimental observations were consistent with the ZV model for the vortex-assisted mechanism of initial dissipation \cite{Zot12prb}. While that work offered micrometer-wide NbN bridges as an alternative to the standard superconducting single-photon detectors, based on nanometer-scale nanowires implemented in long meandering structures, the applicability of the model to superconductors with weaker pinning has lacked experimental scrutiny so far.

Here, we study experimentally the magnetic field dependent photon count rate of a $2\,\mu$m-wide and $10\,\mu$m-long superconducting strip in comparison with a $115$\,nm-wide meander stripline made of MoSi, an important superconducting thin film material widely used in highly efficient superconducting single-photon detectors  \cite{Ver15opt,Cal18apl}.  In addition to weak pinning, MoSi was chosen for the sake of achieving larger experimental critical currents whose values should be close to the critical pair-breaking current in the material, for the high detection efficiency. By carrying out experiments at different values of the perpendicular magnetic field we provide a more direct test of the vortex-antivortex model since the applied magnetic field breaks the translation symmetry for the creation of a vortex-antivortex pair and the current distribution, because the latter is no longer uniform over the cross-section of the bridge. In addition to the wide MoSi strip we also studied the single photon response of a narrow meandering MoSi stripline where an absorbed photon is capable to suppress superconductivity across its whole width (creating the so-called ``hot belt'') and the ``local'' response is not expected.

Our key observation for the wide MoSi strip is that the photon count rate has a crossover current below which the photon count rate (PCR) increases with magnetic field and above which it decreases. Since a similar crossover in the presence of a magnetic field was observed previously in sub-micrometer-wide NbN strips, we believe that the same vortex-assisted mechanism is at work in those narrow strips \cite{Vod15prb}. By contrast, we did not observe a crossover current for the MoSi meander stripline although we reveal a saturation of the PCR at large values of the bias current. This finding implies that in the narrow MoSi strip (whose width is comparable with the width of the NbN strip in \cite{Vod15prb}) the absorbed photon suppresses superconductivity across the whole width of the superconductor that leads to a different dependence of the PCR on the magnetic field.

\section{Experiment}
The MoSi films were deposited by dc magnetron co-sputtering of elemental molybdenum and silicon targets onto Si wafers covered with a $230$\,nm-thick SiO$_2$ layer \cite{Kor14sst}. The atomic percent composition of the Mo$_{68}$Si$_{32}$ films was ensured by using the calibrated deposition rates in conjunction with X-ray reflectivity measurements for deduction of the film thickness. Here we report the results for a straight strip and a meander stripline made of $3.3$\,nm-thick films having a superconducting transition temperature $T_c$ of $3.85$\,K. From the extrapolation of the linear section of upper critical magnetic field $B_{c2}$ near $T_{c}$, we estimate an electron diffusion coefficient of $D$=0.47\,cm$^2$ s$^{-1}$ \cite{Bar16boo} and $B_{c2}(0)$ as 8.66\,T yielding the zero-temperature coherence length $\xi_c(0) \approx 8.7$\,nm. The one spin electronic density of states $N_{0}$ =2.51$\cdot$ 10$^{24}$m$^{3}$K$^{-1}$ was deduced from the Einstein relation. The theoretical depairing current $I_\mathrm{dep}$ at a given temperature was determined from the approximate expression for the critical depairing current, $I_\mathrm{dep}(T)=I_\mathrm{dep}(0)(1-(T/T_c)^2)^{3/2}$, whose applicability for dirty superconductors was justified by Clem and Kogan \cite{Cle12prb}, in agreement with experimental work on aluminium \cite{Rom82prb}.

\begin{figure}[t]
    \includegraphics[width=0.5\textwidth]{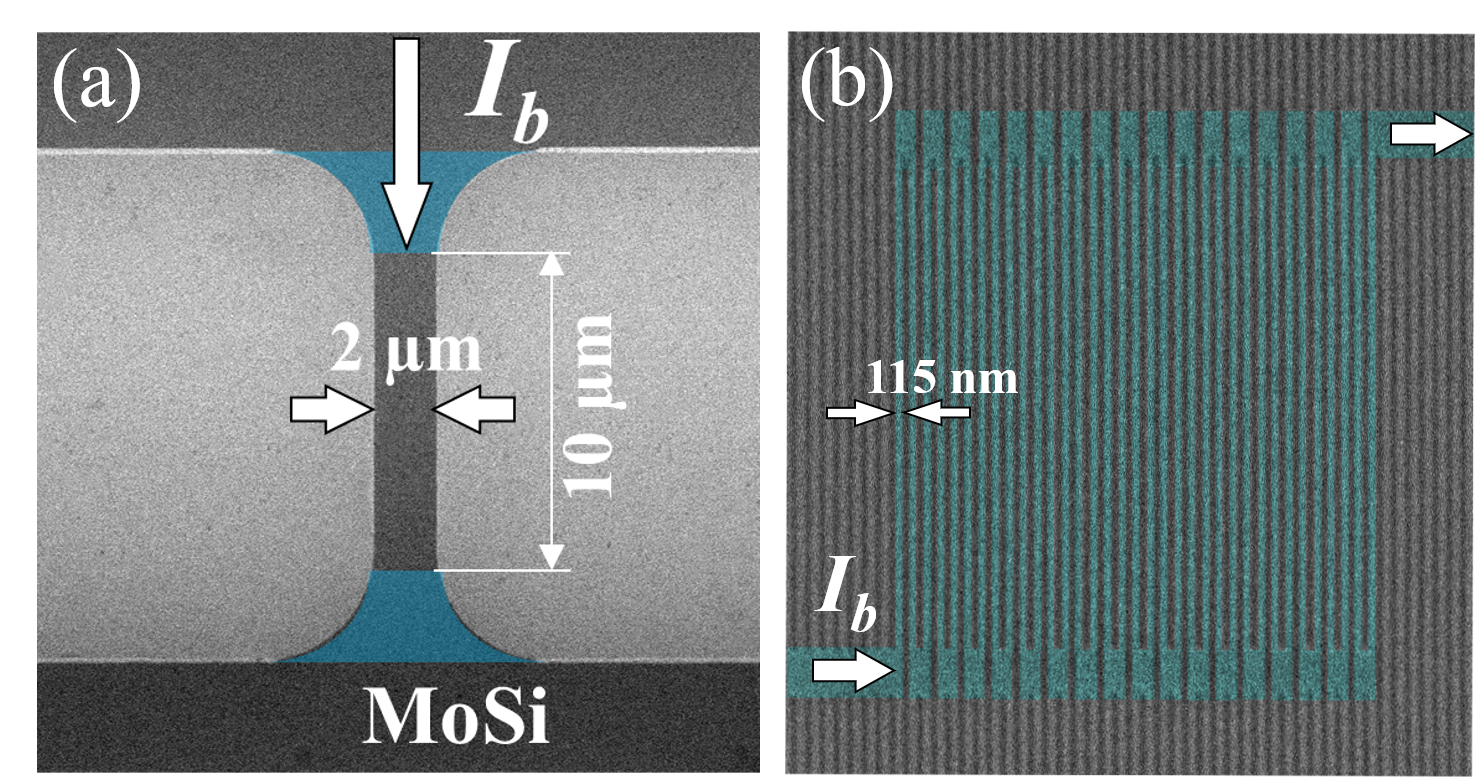}
    \caption{SEM images of the studied samples: (a) $2\,\mathrm\mu$m-wide and $10\,\mathrm\mu$m-long straight strip. The black areas denote the \textalpha-MoSi film and the strip, grey is the etched
    area, and blue denotes the area of MoSi designed to prevent the current-crowding effect. (b) Meander comprised of a $115$\,nm-wide stripline, the total length of the stripline is about $500\,\mu$m.}
    \label{f1}
\end{figure}

Both samples were patterned by using electron beam lithography and reactive ion etching. To prevent current-crowding effects at the sharp strip edges, that may lead to an undesirable reduction of the experimentally measured  critical current, the two ends of the strip were rounded as depicted in Fig.\,\ref{f1}(a). The straight strip is $2\,\mu$m wide and $10\,\mu$m long while the meander stripline is $115$\,nm wide and $500\,\mu$m long. The meander stripline is characterized by a filling factor of 0.58. Both samples are characterized by a magnetic field penetration depth $\lambda(0)$ of  $708$\,nm and a Pearl length $\Lambda(0)$ of $303\,\mu$m, as calculated from their resistivity and $T_c$ values. These estimates are in line with the values deduced from a recent experiments \cite{Zha19arx,Cec19prb} and ensure that $w \ll \Lambda(0)$ for both detectors. Here, $w$ is the strip width. Thus, the current distribution is assumed to be homogeneous across both samples.

To verify that the samples detect single photons, a standard statistical analysis of the photon count rate as a function of the number of incident photons \cite{Kor18pra} was employed. For Poissonian light sources (cw lasers), in the single-photon regime, the PCR is proportional to the photon flux (or incident light power) in the range from $500$\,nm to $1550$\,nm. The setup used for measurements of the PCR and the dark count rate (DCR) in a magnetic field was described elsewhere \cite{Kor18pra}. Light was delivered by a single-mode optical fiber to the sample placed at a distance of $8$\,cm to create a uniformly illuminated field. An external magnetic field, oriented normally to the plane of the sample, was generated by a superconducting solenoid. All measurements described in what follows were done at $T=1.7$\,K. To prevent latching of the wide strip after exceeding the critical current a shunt resistance of $R = 1\,\Omega$ was used. As the shunt is attached to the normal part of the sample a small part of the current flows through the shunt even when the sample is in the superconducting state, which leads to an apparent increase of the critical current in photon counting experiments.

\section{Results and Discussion}
\subsection{Critical currents in magnetic field}

\begin{figure}[t]
    \includegraphics[width=0.50\textwidth]{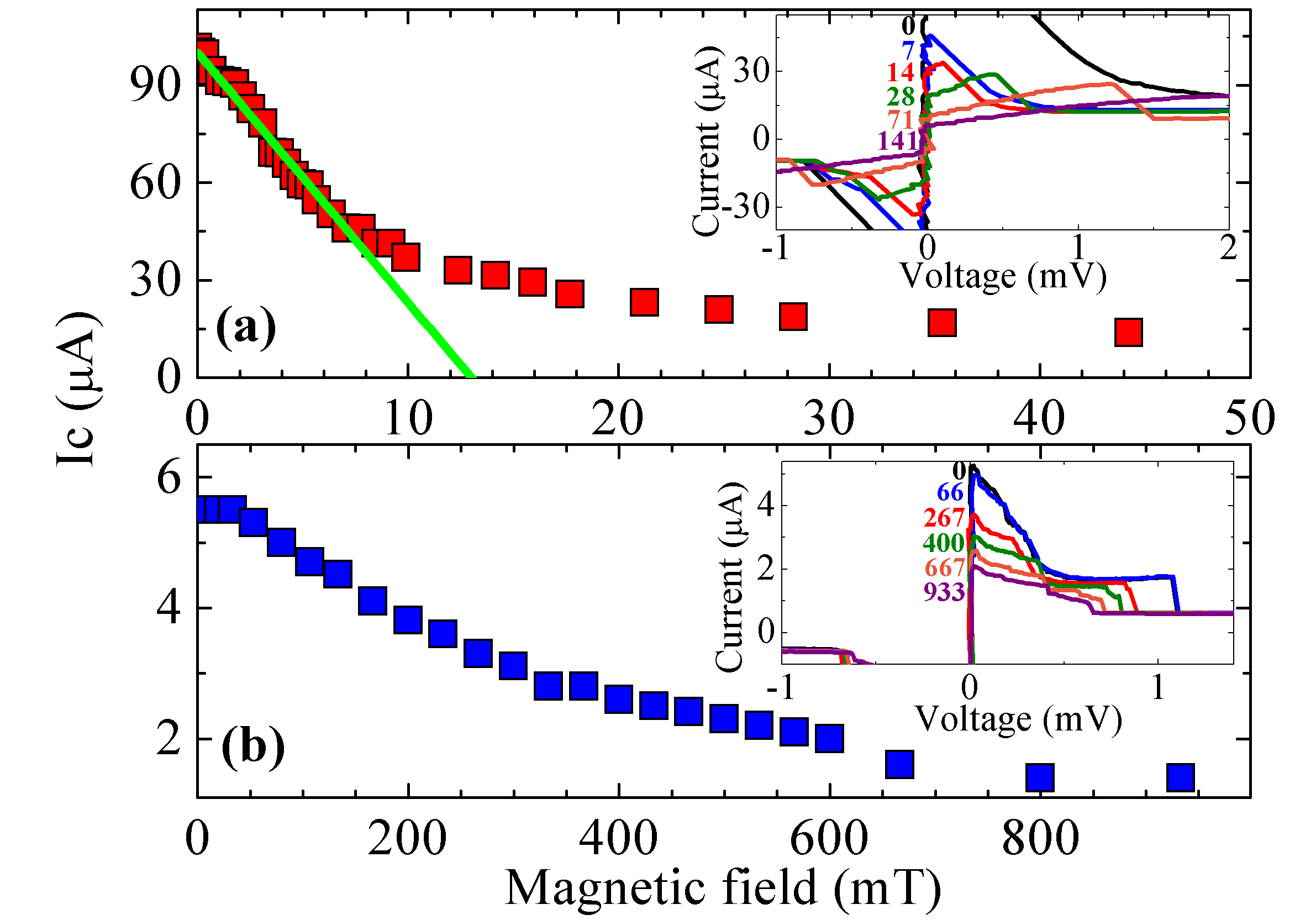}
    \caption{(a) Dependence of the critical current $I_\mathrm{c}(B)$ of the wide MoSi strip on the applied magnetic field. Solid line is a fit to the theoretical $I_\mathrm{c}$ controlled solely by the edge barrier for vortex entry \cite{Plo01prb}. (b) Dependence of the normalized critical current of the MoSi meander. Insets: IV curves of the MoSi strip (top) and the meander (bottom) at different magnetic fields in mT, as labeled close to the curves. For clarity, only IV curves recorded with a negative-to-positive voltage sweep are presented. The IV curves for the wide strip demonstrate a flux-flow regime at $B \gtrsim 10$\,mT.}
    \label{f2}
\end{figure}

Figure~\ref{f2}(a) presents the dependence of the critical current $I_\mathrm{c}$ of the wide strip on the perpendicular magnetic field $B$. Specifically, $I_\mathrm{c}(B)$ exhibits a linear dependence at low fields and its behavior evolves into a more slow nonlinear decrease for fields $B>B^*\simeq 6.9$\,mT. This experimental finding resembles that for $I_\mathrm{c}(B)$ observed for Nb, NbN, and TaN by Lusche \emph{et al}\,\cite{Lus14prb}, Engel \emph{et al}\,\cite{Eng12prb} and Ilin \emph{et al}\,\cite{Ili14prb}, as well as for MoGe by Plourde \emph{et al}\,\cite{Plo01prb}. This crossover from the linear law to the nonlinear dependence can be understood as a transition from $I_\mathrm{c}(B)$ controlled by the edge barrier for vortex penetration at low fields \cite{Kup75pss,Ben98prb,Plo01prb} to $I_\mathrm{c}(B)$ associated with the bulk pinning of vortices at higher fields \cite{Eli02prb}. Note that the nearly linear behavior at $B \lesssim 3$\,mT means that we are still not fully reaching the depairing current at $B=0$, as otherwise the dependence $I_\mathrm{c}(B)$ would, theoretically, be nonlinear, due to the strong pair breaking effect of the depairing current \cite{Mak01epl, And74etp}. Indeed, using the parameters of the sample, for zero magnetic field, we calculate a theoretically expected depairing current density of $j_\mathrm{dep}(B=0)=21\times10^9$\,A/m$^2$. The magnetic field for the suppression of the edge barrier for vortex entry is estimated as $B_\mathrm{s}=\Phi_0/2\sqrt{3}\pi \xi W =11$\,mT. Both estimates are not far from but not very close to the experimental values $j_\mathrm{c}(B=0)=17\times10^9$\,A/m$^2$ (corresponding to $I_\mathrm{c}(B=0)=110\,\mu$A) and $B^*=B_\mathrm{s}/2=6.5$\,mT.

We like to point out that in our previous work on MoSi \cite{Kor14sst} the critical current density at $T$=1.7\,K in zero magnetic field was in the range from $15\times10^9$\,A/m$^2$ to $18\times10^9$\,A/m$^2$, which is about 45\% to 52\% of the depairing critical current density. In that experiment we used meandering strips with width $114$\,nm. In the present study, we reach critical current densities of about $70\%$ of the depairing current density for the bridge, and $64\%$ for the meander stripline.

Importantly, the dependence $I_\mathrm{c}(B)$ for the MoSi meander does not exhibit a linear behavior at low magnetic fields, referring to Fig. \ref{f2}(b), although $I_\mathrm{c}(0)\simeq 5.5\,\mu$A is about 64$\%$ of the theoretical depairing current. We believe that it is related to the presence of a defect somewhere inside the meander stripline, which may result in a rounding of $I_\mathrm{c}(B)$ at low fields even if there is an edge barrier for vortex entry (see for example Fig. 6 in Ref. \cite{Vod15sst}). Remarkably, for the MoSi meander we did not observe a flux-flow branch in the current-voltage characteristics (IV curve) up to $B=933$\,mT, see the inset in Fig. \ref{f2}(b), which is much larger than the theoretical estimate $B^*=95$ \,mT. By contrast, for the MoSi strip the flux flow branch in the IV curve appears approximately at $B \gtrsim B^*$, see the inset in Fig. \ref{f2}(a), and its evolution with increasing $B$ is similar to that observed by Samoilov \textit{et al} \cite{Sam95prl} for Mo$_3$Si superconducting strips.

\subsection{Count rates in magnetic field}
\begin{figure}[t]
\includegraphics[width=0.50\textwidth]{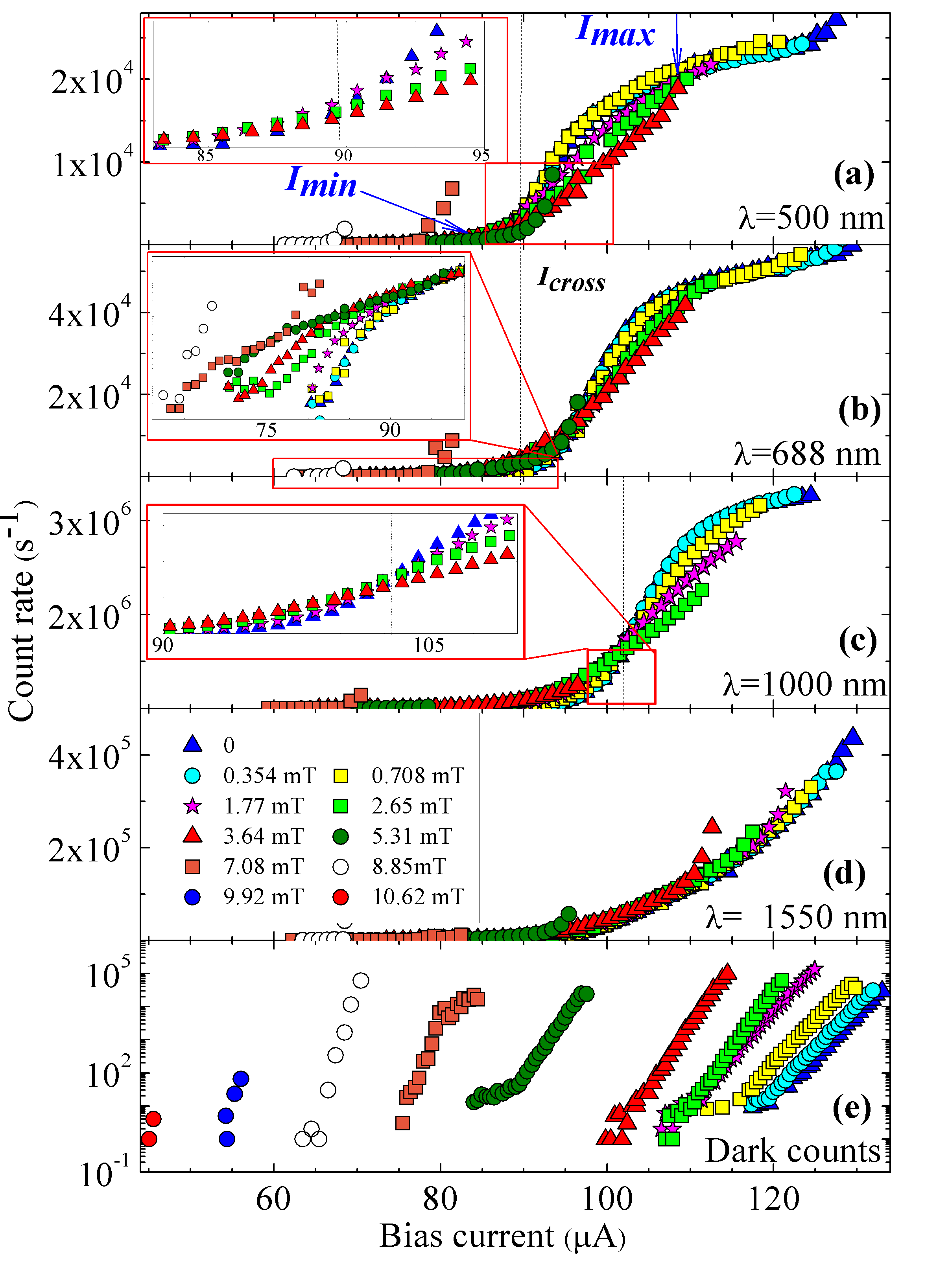}
    \caption{(a-d) Dependence of the PCR of the wide strip on the current at different magnetic fields and for different wavelengths of incident photons. At $B=0$ the count rate saturates at a relatively large current except for photons with $\lambda=1550$\,nm. One can see the presence of a crossover current $I_\mathrm{cross}$ for wavelengths $500 -1000$\,nm  (panels (a)-(c)), above which the count rate decreases, while for currents $I<I_\mathrm{cross}$, the PCR increases with increasing magnetic field. In panel (a) the arrows mark the current $I_\mathrm{min}$, at which the PCR starts to grow sharply and $I_\mathrm{max}$, at which it is almost saturated. The insets in panels (a, linear scale), (b, logarithmic scale) and (c, linear scale) show enlarged parts of the regions marked by the red rectangles in semi-log scale. (e) Dependence of the DCR on the bias current at different $B$ values. At $B>10$\,mT dark counts are practically absent.}
    \label{f3}
\end{figure}

Figure~\ref{f3} presents the PCR and the DCR of the wide strip as a function of the bias current and the applied magnetic field for different photon wavelengths. The PCR exhibits a tendency to saturate as $I\to I_\mathrm{c}(B=0)$ (except for $\lambda =1550$\,nm). The saturation means that we approach the $100 \%$ intrinsic detection efficiency. In our previous work on micron-wide NbN bridges~\cite{Kor18pra} the saturation of the photon count rate was only observed on a logarithmic scale, and it was tentatively explained by the shape of the used constriction-type microbridges. In the present case, the not fully complete saturation on the linear scale is most likely still connected to the presence of tapered active areas near the ends of the bridge which are also exposed to the photon flux. These areas are indicated in blue in the SEM image in Fig.~\ref{f1}(a).

In the applied magnetic field the PCR changes gradually as a function of the bias current. Specifically, at small currents the number of counts increases with increasing magnetic field, whereas at larger currents the PCR decreases with increasing magnetic field. At intermediated currents, we define a crossover current $I_\mathrm{cross}$, as shown in Fig. \ref{f3}(a)-(c), at which the PCR is independent of the magnetic field. The crossover current is only observed for wavelengths at which the PCR gets saturated, as is indicative of the saturation of the detection efficiency. This peculiar behavior and the presence of the crossover current is a fingerprint of the vortex-antivortex assisted photon counting mechanism and the ``locality'' of the photoresponse in the wide MoSi strip.

The evolution of the DCR$(I)$ with increasing magnetic field is presented in Fig. \ref{f3}(e). The dependence DCR($I$) is similar to that obtained for nanoscale meander-type NbN detectors consisting of $35$ connected strips, each having a width of about $100$\,nm, and a length of $7\,\mu$m \cite{Vod15prb}. The origin of the dark counts is assumed to be due to the occasional entry of vortices due to random processes. According to the London model, at $I\lesssim I_\mathrm{c}$ the amplitude of the energy barrier $\delta E$ varies as $\delta E \sim 1-I/I_\mathrm{c}$ \cite{Bar10prb}. As the field increases, the linear dependence of $\mathrm{log(DCR)}$ on the current shifts towards lower currents since $I_\mathrm{c}$ decreases with increasing $B$ (see also Fig.~\ref{f2}). Dark counts are practically absent at fields $B>10$\,mT, that agrees well with the field at which a low-resistive branch appears in the IV curve in the inset of Fig.~\ref{f2}(a), identified as a flux-flow regime. At such a low current $I\sim I_\mathrm{c}$ the passage of a single vortex across the strip is not enough to heat the sample and, hence, dark counts no longer appear.

\begin{figure}[t]
    \includegraphics[width=0.5\textwidth]{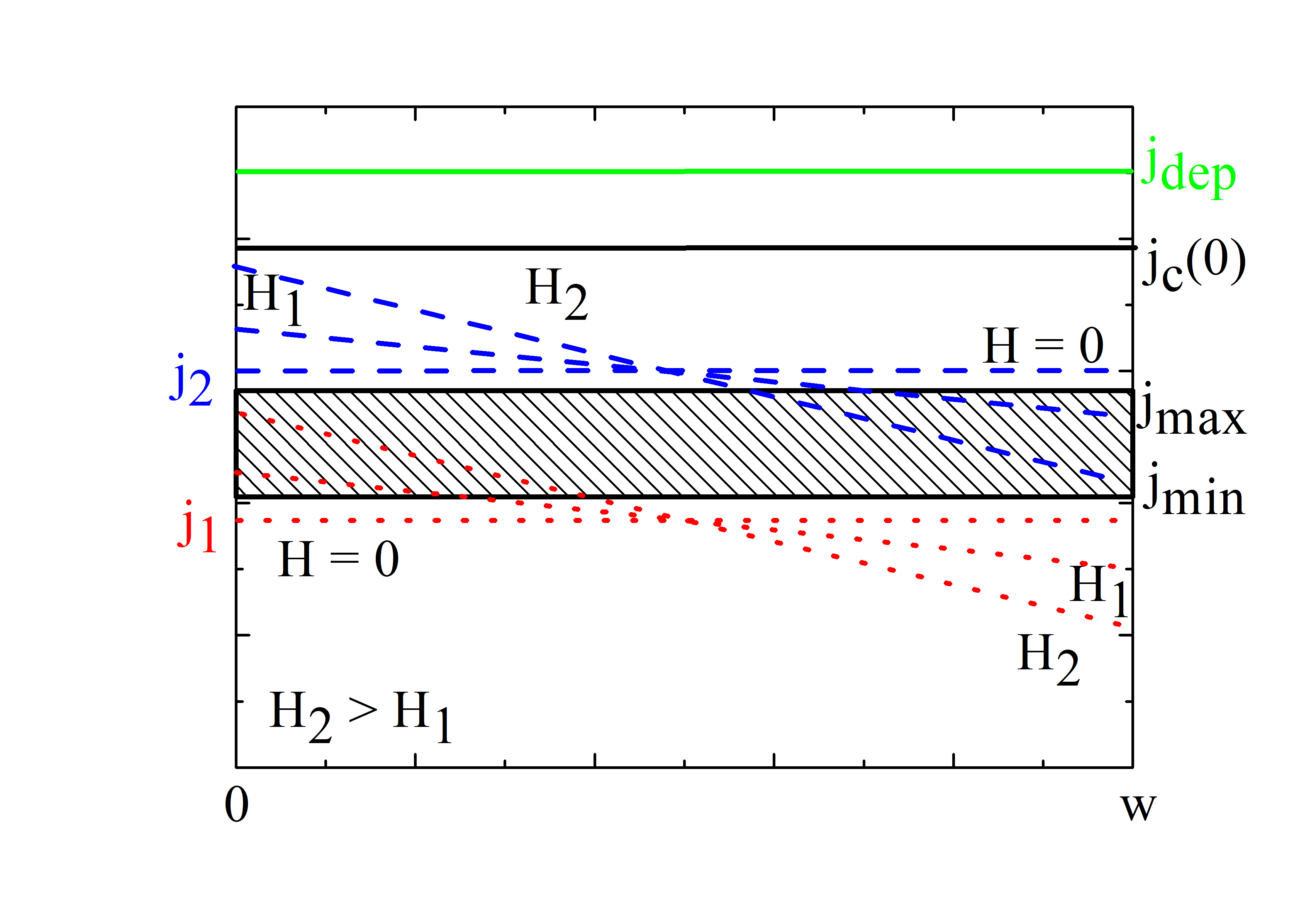}
    \caption{Graphical representation of the current density across the width of the superconducting strip. The vertical axis represents values of the current density. The upper green line represents the theoretical depairing current density $j_\mathrm{dep}=I_\mathrm{dep}/wd$, while $j_\mathrm{c}(0)=I_\mathrm{c}(B=0)/wd$ represents the experimentally deduced critical current density. The range of currents between $I_\mathrm{min}=j_\mathrm{min}wd$ and $I_\mathrm{max}=j_\mathrm{max}wd$ represents the band over which the photon count rate increases nonlinearly from an almost zero value up to the maximal one (see Fig. \ref{f3}). At the current $I_2=j_2wd>I_\mathrm{max}$ and $H=0$, the photon count rate is equal to the maximal value PCR$_\mathrm{max}$. At the current $I_1=j_1wd<I_\mathrm{min}$ and $B=0$, the PCR is equal to zero. An applied magnetic field tilts the current density distribution, leading to a non-uniform response to photons. At $x=0$ PCR gets saturated, where as $x=w$ it increases from zero to a finite value. Obviously, the reason is that a perpendicular field adds a current to the bias current at $x=0$ and subtracts it at $x=w$.}
    \label{f4}
\end{figure}

To clarify the origin of the crossover current and the dependence of $\mathrm{PCR}(I)$ on the magnetic field, in Fig. \ref{f4} we present a sketch of the current distribution in a superconducting strip at different values of currents and magnetic fields. Thus, when $B=0$ the current is distributed uniformly, as depicted by horizontal lines. In a wide strip the value of the calculated detection current (defined as a current at which the superconductor switches to a resistive state after absorption of a photon) depends weakly on the coordinate of the incident photon across the strip, except for the areas near the edges with a width of about the diameter of the photon-induced hot spot. This, in particularly, contrasts with narrow strips $w \gtrsim D_\mathrm{HS}$ in which the detection current varies strongly with the coordinate \cite{Zot14sst,Eng15tas,Ren15nal}. The uniformity of the current distribution is a consequence of the ``locality'' of the photoresponse in strips with $w \gg D_{HS}$. In the theoretical model, it results in an almost step-like dependence of the photon counting rate on current, which is not observed in the experiment. This difference between experiment and theory may be due to the presence of unavoidable electronic inhomogeneities in the strip or due to Fano fluctuations, i.e. fluctuations in the number of non-equilibrium quasiparticles produced by the incident photon \cite{Koz17prb}. Both mechanisms lead to a smeared dependence of the photon count rate on current. To take this observed ``fuzziness'' into account, in Fig. \ref{f4} we draw a minimal current $I_\mathrm{min}=j_\mathrm{min}wd$ and a maximal current $I_\mathrm{max}=j_\mathrm{max}wd$, which are also depicted in Fig. \ref{f3}. In our simple model, we assume that the probability for the photon to be detected increases smoothly from zero at $I=I_\mathrm{min}$, where the current density is equal to $j_\mathrm{min}$, up to unity where the current $I=I_\mathrm{max}$ (and the current density is $j=j_\mathrm{max}$). So we have a band of bias current values over which the photon count rate rises from close to zero to close to 100\%.

When a perpendicular magnetic field is applied, the supercurrent density distribution becomes nonuniform. Namely, at one edge the current is going out of the plane and at the other edge into the plane. It means that at one edge the current density increases whereas at the other edge it decreases (see Fig. \ref{f4}). If the original bias current is larger than $I_\mathrm{max}$, in the part of the strip where the current density drops below $j_\mathrm{max}$ the PCR will decrease. If the current is smaller than $I_\mathrm{min}$ in the part of the strip with the current density is larger than $I_\mathrm{min}$ the PCR increases. Between $I_\mathrm{max}$ and $I_\mathrm{min}$ there is a crossover current at which the PCR does not change with magnetic field. Indeed, such a crossover is observed experimentally for $\lambda = 500, 688, 1000$ nm. The fact that such a crossover current is not seen for infrared photons with $\lambda = 1550$ nm is most likely related to the absence of saturation in the dependence of $\mathrm{PCR}(I)$.

\begin{figure}[t]
    \includegraphics[width=0.5\textwidth]{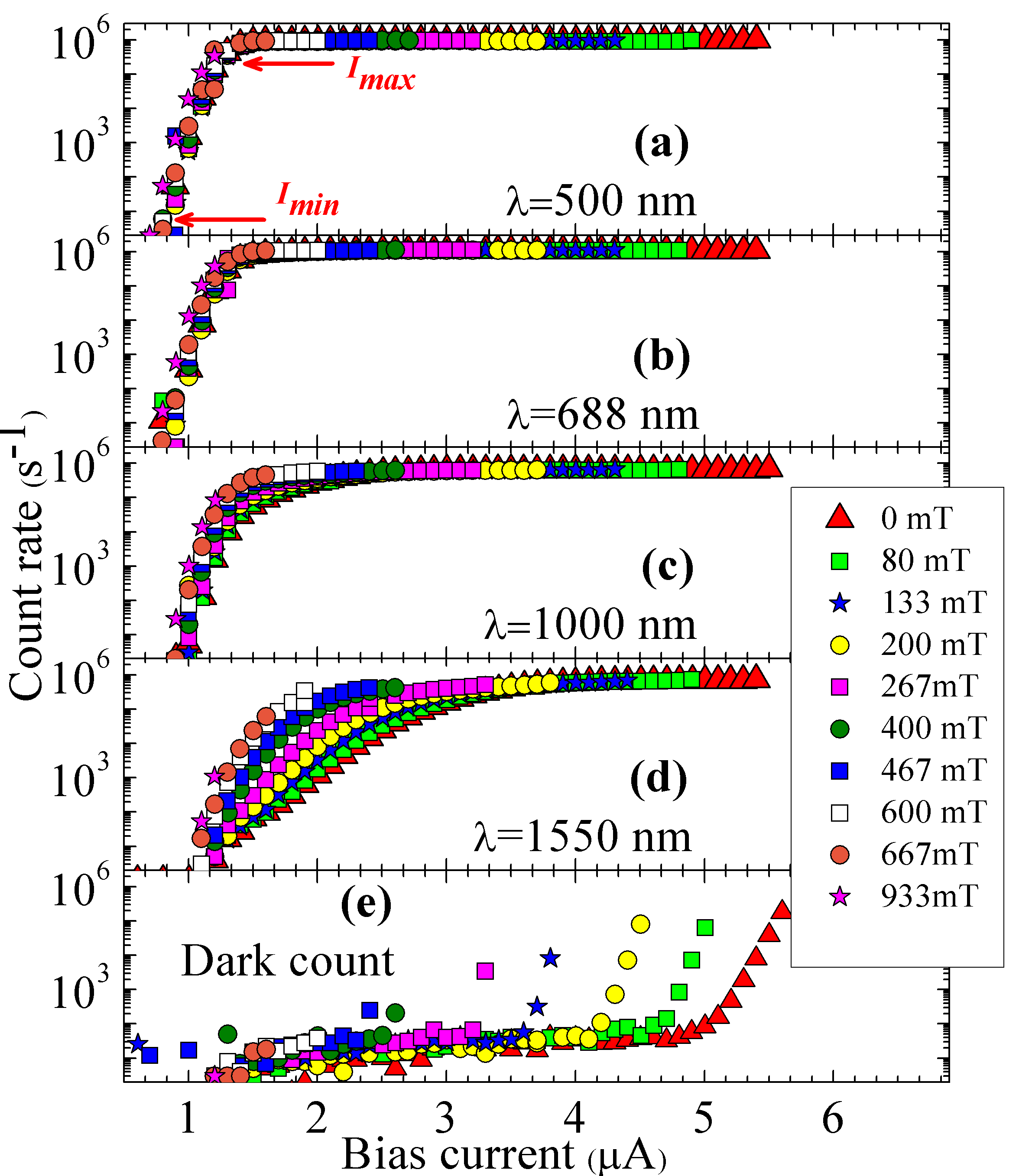}
    \caption{PCR vs bias current in different magnetic fields measured for the MoSi meander made of a $115$\,nm-wide stripline. Panels (a)-(d) present counts for the indicated wavelengths, panel (e) depicts dark count rates. The currents $I_\mathrm{min}$ and $I_\mathrm{max}$, as in Fig.\,\ref{f3}, indicate currents at which the photon count rate starts to grow sharply and saturates, respectively. In contradistinction with Fig.\,\ref{f3}, here the data are plotted on a logarithmic scale to emphasize the dependence on magnetic field, which is hardly visible on a linear scale.}
    \label{f5}
\end{figure}

In contrast to the wide strip, no crossover current is observed for the MoSi meander, see Fig. \ref{f5}, and we also refer to a previous work on narrow MoSi meanders \cite{Kor15tas} where no crossover currents were observed as well. Note that for studied wavelengths, single photon counting starts at approximately the same current $\sim 1\,\mu$A which is close to the retrapping current of the MoSi meander $I_\mathrm{r} \simeq 0.6\,\mu$A (see the inset in Fig. \ref{f2}(b)). This fact in conjunction with the found dependence of the PCR on the magnetic field allows us to suppose that the absorbed photon fully or partially suppresses superconductivity across the whole stripline, forming a hot belt. Indeed, in this case the applied magnetic field should decrease the detection current (which does not depend on coordinate) as it does with the critical current (see Fig. \ref{f2}(b)). Therefore the dependence $\mathrm{PCR}(I)$ should shift towards lower currents as it follows from Refs. \cite{Zot14sst,Bul12prb} and resemble the evolution of $\mathrm{DCR}(I)$ with increasing $B$. However, since the minimal detection current cannot exceed the retrapping current, as at $I<I_\mathrm{r}$ the growing normal domain cannot appear in the superconducting strip \cite{Vod17pra}, this shift is partially observed only for $\lambda=1550$\,nm. Note, that in the wide MoSi strip $I_\mathrm{r} =18\,\mu$A (see the inset in Fig. \ref{f2}(a)) and the PCR strongly depends on magnetic field for all wavelengths.

The crossover current was earlier observed in our previous work on NbN meanders~\cite{Vod15prb} having almost the same stripline width as the MoSi meander studied here. Because the critical temperature and the thickness of the stripline in the NbN meander were larger than the respective parameters for the MoSi one, the diameter of a hot (normal) spot in NbN should be about three times smaller, if one uses the energy conservation law for its estimation (see Eq. (38) in \cite{Vod17pra}). Therefore we believe that the ``local'' model and the vortex-antivortex detection mechanism should stipulate the photon detection in those NbN meanders too. Because $I_\mathrm{min}$ is close to the retrapping current of the NbN strip at $\lambda = 450$\,nm, the PCR weakly depends on magnetic field there (see Fig. \ref{f2}(d) in Ref. \cite{Vod15prb}) like in our MoSi meander.

Finally, to complete the discussion of the obtained results, we would like to note that the effect of the perpendicular magnetic field on the DCR and the PCR were also studied by Engel \textit{et al} \cite{Eng12prb} and Lusche \textit{et al} \cite{Lus14prb} for TaN and NbN meanders with similar widths of the stripline. In Ref. \cite{Eng12prb} no effect of magnetic field on the PCR was observed at fields up to $10$\,mT while the DCR grew as the field value increased. In our MoSi meander the PCR is also changed much weaker in comparison with the DCR at a fixed current (see Fig. \ref{f5}), which we attribute to the non-fluctuation nature of the PCR in contrast with the DCR. To find a decrease of the PCR and a crossover current one needs to be in the regime of the PCR saturation which is indicative of reaching the $100 \%$ intrinsic detection efficiency, and the width of the superconducting strip should be larger than the diameter of the photon-induced hot spot for applicability of the ``local'' model. In Ref. \cite{Lus14prb}, the former condition was not fulfilled and only an increase of the PCR was experimentally observed.

\section{Conclusions}
To summarize, we have studied the photon count rate in a wide strip and a meander with narrow stripline made of superconducting MoSi. We have found different dependences of the photon count rate on the perpendicular magnetic field in these two systems. Namely, a crossover current $I_\textrm{cross}$ has been observed for the wide strip, below which the count rate increases with increasing magnetic field and above which it decreases. This observation is in agreement with the photon-induced vortex generation model, and it is explained by the magnetic field induced screening currents in the strip. By contrast, in the narrow MoSi meander no crossover current has been observed and the dependence of the photon count rate on the magnetic field has been revealed to be much weaker. We explain this finding by a low value of the detection current, which is close to the retrapping current of the meander and the photon-induced suppression of superconductivity across the whole width of the meander stripline.

\section*{Acknowledgments}
We thank Teunis Klapwijk for stimulating discussions, and Gregory Goltsman for his stimulating interest in this work and his encouragement, as well as for the creation of the used laboratory infrastructure. This work was supported by the Russian Science Foundation (RSF) grant No. 17-72-30036 (in the part concerning experimental study of single photon response of MoSi meander and strip). Research work of MYM and OVD  was partially conducted within the framework of the COST Action CA16218 (NANOCOHYBRI) of the European Cooperation in Science and Technology (thin film deposition, structural and superconducting properties of thin films). DV also acknowledges support by the Russian Foundation for Basic Research (RFBR), grant No. 18-29-20100 (in the part concerning theoretical analysis of experimental results).
\input Korneeva_PRB_submission.bbl

\end{document}

%% file: Korneeva_PRB_submission.bbl
%

%% file: Korneeva_PRB_submission.bbl
\begin{thebibliography}{32}%
\makeatletter
\providecommand \@ifxundefined [1]{%
 \@ifx{#1\undefined}
}%
\providecommand \@ifnum [1]{%
 \ifnum #1\expandafter \@firstoftwo
 \else \expandafter \@secondoftwo
 \fi
}%
\providecommand \@ifx [1]{%
 \ifx #1\expandafter \@firstoftwo
 \else \expandafter \@secondoftwo
 \fi
}%
\providecommand \natexlab [1]{#1}%
\providecommand \enquote  [1]{``#1''}%
\providecommand \bibnamefont  [1]{#1}%
\providecommand \bibfnamefont [1]{#1}%
\providecommand \citenamefont [1]{#1}%
\providecommand \href@noop [0]{\@secondoftwo}%
\providecommand \href [0]{\begingroup \@sanitize@url \@href}%
\providecommand \@href[1]{\@@startlink{#1}\@@href}%
\providecommand \@@href[1]{\endgroup#1\@@endlink}%
\providecommand \@sanitize@url [0]{\catcode `\\12\catcode `\$12\catcode
  `\&12\catcode `\#12\catcode `\^12\catcode `\_12\catcode `\%12\relax}%
\providecommand \@@startlink[1]{}%
\providecommand \@@endlink[0]{}%
\providecommand \url  [0]{\begingroup\@sanitize@url \@url }%
\providecommand \@url [1]{\endgroup\@href {#1}{\urlprefix }}%
\providecommand \urlprefix  [0]{URL }%
\providecommand \Eprint [0]{\href }%
\providecommand \doibase [0]{https://doi.org/}%
\providecommand \selectlanguage [0]{\@gobble}%
\providecommand \bibinfo  [0]{\@secondoftwo}%
\providecommand \bibfield  [0]{\@secondoftwo}%
\providecommand \translation [1]{[#1]}%
\providecommand \BibitemOpen [0]{}%
\providecommand \bibitemStop [0]{}%
\providecommand \bibitemNoStop [0]{.\EOS\space}%
\providecommand \EOS [0]{\spacefactor3000\relax}%
\providecommand \BibitemShut  [1]{\csname bibitem#1\endcsname}%
\let\auto@bib@innerbib\@empty
\bibitem [{\citenamefont {Gol'tsman}\ \emph {et~al.}(2001)\citenamefont
  {Gol'tsman}, \citenamefont {Okunev}, \citenamefont {Chulkova}, \citenamefont
  {Lipatov}, \citenamefont {Semenov}, \citenamefont {Smirnov}, \citenamefont
  {Voronov}, \citenamefont {Dzardanov}, \citenamefont {Williams},\ and\
  \citenamefont {Sobolewski}}]{Gol01apl}%
  \BibitemOpen
  \bibfield  {author} {\bibinfo {author} {\bibfnamefont {G.~N.}\ \bibnamefont
  {Gol'tsman}}, \bibinfo {author} {\bibfnamefont {O.}~\bibnamefont {Okunev}},
  \bibinfo {author} {\bibfnamefont {G.}~\bibnamefont {Chulkova}}, \bibinfo
  {author} {\bibfnamefont {A.}~\bibnamefont {Lipatov}}, \bibinfo {author}
  {\bibfnamefont {A.}~\bibnamefont {Semenov}}, \bibinfo {author} {\bibfnamefont
  {K.}~\bibnamefont {Smirnov}}, \bibinfo {author} {\bibfnamefont
  {B.}~\bibnamefont {Voronov}}, \bibinfo {author} {\bibfnamefont
  {A.}~\bibnamefont {Dzardanov}}, \bibinfo {author} {\bibfnamefont
  {C.}~\bibnamefont {Williams}},\ and\ \bibinfo {author} {\bibfnamefont
  {R.}~\bibnamefont {Sobolewski}},\ }\bibfield  {title} {\bibinfo {title}
  {Picosecond superconducting single-photon optical detector},\ }\href
  {https://doi.org/10.1063/1.1388868} {\bibfield  {journal} {\bibinfo
  {journal} {Appl. Phys. Lett.}\ }\textbf {\bibinfo {volume} {79}},\ \bibinfo
  {pages} {705} (\bibinfo {year} {2001})}\BibitemShut {NoStop}%
\bibitem [{\citenamefont {Natarajan}\ \emph {et~al.}(2012)\citenamefont
  {Natarajan}, \citenamefont {Tanner},\ and\ \citenamefont
  {Hadfield}}]{Nat12sst}%
  \BibitemOpen
  \bibfield  {author} {\bibinfo {author} {\bibfnamefont {C.~M.}\ \bibnamefont
  {Natarajan}}, \bibinfo {author} {\bibfnamefont {M.~G.}\ \bibnamefont
  {Tanner}},\ and\ \bibinfo {author} {\bibfnamefont {R.~H.}\ \bibnamefont
  {Hadfield}},\ }\bibfield  {title} {\bibinfo {title} {Superconducting nanowire
  single-photon detectors: physics and applications},\ }\href
  {https://doi.org/10.1088/0953-2048/25/6/063001} {\bibfield  {journal}
  {\bibinfo  {journal} {Supercond. Sci. Technol.}\ }\textbf {\bibinfo {volume}
  {25}},\ \bibinfo {pages} {063001} (\bibinfo {year} {2012})}\BibitemShut
  {NoStop}%
\bibitem [{\citenamefont {Korneeva}\ \emph {et~al.}(2018)\citenamefont
  {Korneeva}, \citenamefont {Vodolazov}, \citenamefont {Semenov}, \citenamefont
  {Florya}, \citenamefont {Simonov}, \citenamefont {Baeva}, \citenamefont
  {Korneev}, \citenamefont {Goltsman},\ and\ \citenamefont
  {Klapwijk}}]{Kor18pra}%
  \BibitemOpen
  \bibfield  {author} {\bibinfo {author} {\bibfnamefont {Y.~P.}\ \bibnamefont
  {Korneeva}}, \bibinfo {author} {\bibfnamefont {D.~Y.}\ \bibnamefont
  {Vodolazov}}, \bibinfo {author} {\bibfnamefont {A.~V.}\ \bibnamefont
  {Semenov}}, \bibinfo {author} {\bibfnamefont {I.~N.}\ \bibnamefont {Florya}},
  \bibinfo {author} {\bibfnamefont {N.}~\bibnamefont {Simonov}}, \bibinfo
  {author} {\bibfnamefont {E.}~\bibnamefont {Baeva}}, \bibinfo {author}
  {\bibfnamefont {A.~A.}\ \bibnamefont {Korneev}}, \bibinfo {author}
  {\bibfnamefont {G.~N.}\ \bibnamefont {Goltsman}},\ and\ \bibinfo {author}
  {\bibfnamefont {T.~M.}\ \bibnamefont {Klapwijk}},\ }\bibfield  {title}
  {\bibinfo {title} {Optical single-photon detection in micrometer-scale {NbN}
  bridges},\ }\href {https://doi.org/10.1103/PhysRevApplied.9.064037}
  {\bibfield  {journal} {\bibinfo  {journal} {Phys. Rev. Appl.}\ }\textbf
  {\bibinfo {volume} {9}},\ \bibinfo {pages} {064037} (\bibinfo {year}
  {2018})}\BibitemShut {NoStop}%
\bibitem [{\citenamefont {Zotova}\ and\ \citenamefont
  {Vodolazov}(2012)}]{Zot12prb}%
  \BibitemOpen
  \bibfield  {author} {\bibinfo {author} {\bibfnamefont {A.~N.}\ \bibnamefont
  {Zotova}}\ and\ \bibinfo {author} {\bibfnamefont {D.~Y.}\ \bibnamefont
  {Vodolazov}},\ }\bibfield  {title} {\bibinfo {title} {Photon detection by
  current-carrying superconducting film: A time-dependent {Ginzburg-Landau}
  approach},\ }\href {https://doi.org/10.1103/PhysRevB.85.024509} {\bibfield
  {journal} {\bibinfo  {journal} {Phys. Rev. B}\ }\textbf {\bibinfo {volume}
  {85}},\ \bibinfo {pages} {024509} (\bibinfo {year} {2012})}\BibitemShut
  {NoStop}%
\bibitem [{\citenamefont {Vodolazov}(2017)}]{Vod17pra}%
  \BibitemOpen
  \bibfield  {author} {\bibinfo {author} {\bibfnamefont {D.~Y.}\ \bibnamefont
  {Vodolazov}},\ }\bibfield  {title} {\bibinfo {title} {Single-photon detection
  by a dirty current-carrying superconducting strip based on the
  kinetic-equation approach},\ }\href
  {https://doi.org/10.1103/PhysRevApplied.7.034014} {\bibfield  {journal}
  {\bibinfo  {journal} {Phys. Rev. Appl.}\ }\textbf {\bibinfo {volume} {7}},\
  \bibinfo {pages} {034014} (\bibinfo {year} {2017})}\BibitemShut {NoStop}%
\bibitem [{\citenamefont {Verma}\ \emph {et~al.}(2015)\citenamefont {Verma},
  \citenamefont {Korzh}, \citenamefont {Bussi{\`e}res}, \citenamefont
  {Horansky}, \citenamefont {Dyer}, \citenamefont {Lita}, \citenamefont
  {Vayshenker}, \citenamefont {Marsili}, \citenamefont {Shaw}, \citenamefont
  {Zbinden}, \citenamefont {Mirin},\ and\ \citenamefont {W.}}]{Ver15opt}%
  \BibitemOpen
  \bibfield  {author} {\bibinfo {author} {\bibfnamefont {V.~B.}\ \bibnamefont
  {Verma}}, \bibinfo {author} {\bibfnamefont {B.}~\bibnamefont {Korzh}},
  \bibinfo {author} {\bibfnamefont {F.}~\bibnamefont {Bussi{\`e}res}}, \bibinfo
  {author} {\bibfnamefont {R.~D.}\ \bibnamefont {Horansky}}, \bibinfo {author}
  {\bibfnamefont {S.~D.}\ \bibnamefont {Dyer}}, \bibinfo {author}
  {\bibfnamefont {A.~E.}\ \bibnamefont {Lita}}, \bibinfo {author}
  {\bibfnamefont {I.}~\bibnamefont {Vayshenker}}, \bibinfo {author}
  {\bibfnamefont {F.}~\bibnamefont {Marsili}}, \bibinfo {author} {\bibfnamefont
  {M.~D.}\ \bibnamefont {Shaw}}, \bibinfo {author} {\bibfnamefont
  {H.}~\bibnamefont {Zbinden}}, \bibinfo {author} {\bibfnamefont {R.~P.}\
  \bibnamefont {Mirin}},\ and\ \bibinfo {author} {\bibfnamefont {N.~S.}\
  \bibnamefont {W.}},\ }\bibfield  {title} {\bibinfo {title} {High-efficiency
  superconducting nanowire single-photon detectors fabricated from mosi
  thin-films},\ }\href {https://doi.org/10.1364/OE.23.033792} {\bibfield
  {journal} {\bibinfo  {journal} {Optics Express}\ }\textbf {\bibinfo {volume}
  {23}},\ \bibinfo {pages} {33792} (\bibinfo {year} {2015})}\BibitemShut
  {NoStop}%
\bibitem [{\citenamefont {Caloz}\ \emph {et~al.}(2018)\citenamefont {Caloz},
  \citenamefont {Perrenoud}, \citenamefont {Autebert}, \citenamefont {Korzh},
  \citenamefont {Weiss}, \citenamefont {Sch{\"o}nenberger}, \citenamefont
  {Warburton}, \citenamefont {Zbinden},\ and\ \citenamefont
  {Bussi{\`e}res}}]{Cal18apl}%
  \BibitemOpen
  \bibfield  {author} {\bibinfo {author} {\bibfnamefont {M.}~\bibnamefont
  {Caloz}}, \bibinfo {author} {\bibfnamefont {M.}~\bibnamefont {Perrenoud}},
  \bibinfo {author} {\bibfnamefont {C.}~\bibnamefont {Autebert}}, \bibinfo
  {author} {\bibfnamefont {B.}~\bibnamefont {Korzh}}, \bibinfo {author}
  {\bibfnamefont {M.}~\bibnamefont {Weiss}}, \bibinfo {author} {\bibfnamefont
  {C.}~\bibnamefont {Sch{\"o}nenberger}}, \bibinfo {author} {\bibfnamefont
  {R.~J.}\ \bibnamefont {Warburton}}, \bibinfo {author} {\bibfnamefont
  {H.}~\bibnamefont {Zbinden}},\ and\ \bibinfo {author} {\bibfnamefont
  {F.}~\bibnamefont {Bussi{\`e}res}},\ }\bibfield  {title} {\bibinfo {title}
  {High-detection efficiency and low-timing jitter with amorphous
  superconducting nanowire single-photon detectors},\ }\href
  {https://doi.org/10.1063/1.5010102} {\bibfield  {journal} {\bibinfo
  {journal} {Applied Physics Letters}\ }\textbf {\bibinfo {volume} {112}},\
  \bibinfo {pages} {061103} (\bibinfo {year} {2018})}\BibitemShut {NoStop}%
\bibitem [{\citenamefont {Vodolazov}\ \emph
  {et~al.}(2015{\natexlab{a}})\citenamefont {Vodolazov}, \citenamefont
  {Korneeva}, \citenamefont {Semenov}, \citenamefont {Korneev},\ and\
  \citenamefont {Goltsman}}]{Vod15prb}%
  \BibitemOpen
  \bibfield  {author} {\bibinfo {author} {\bibfnamefont {D.~Y.}\ \bibnamefont
  {Vodolazov}}, \bibinfo {author} {\bibfnamefont {Y.~P.}\ \bibnamefont
  {Korneeva}}, \bibinfo {author} {\bibfnamefont {A.~V.}\ \bibnamefont
  {Semenov}}, \bibinfo {author} {\bibfnamefont {A.~A.}\ \bibnamefont
  {Korneev}},\ and\ \bibinfo {author} {\bibfnamefont {G.~N.}\ \bibnamefont
  {Goltsman}},\ }\bibfield  {title} {\bibinfo {title} {Vortex-assisted
  mechanism of photon counting in a superconducting nanowire single-photon
  detector revealed by external magnetic field},\ }\href
  {https://doi.org/10.1103/PhysRevB.92.104503} {\bibfield  {journal} {\bibinfo
  {journal} {Phys. Rev. B}\ }\textbf {\bibinfo {volume} {92}},\ \bibinfo
  {pages} {104503} (\bibinfo {year} {2015}{\natexlab{a}})}\BibitemShut
  {NoStop}%
\bibitem [{\citenamefont {Korneeva}\ \emph {et~al.}(2014)\citenamefont
  {Korneeva}, \citenamefont {Mikhailov}, \citenamefont {Pershin}, \citenamefont
  {Manova}, \citenamefont {Divochiy}, \citenamefont {Vakhtomin}, \citenamefont
  {Korneev}, \citenamefont {Smirnov}, \citenamefont {Sivakov}, \citenamefont
  {Devizenko},\ and\ \citenamefont {Goltsman}}]{Kor14sst}%
  \BibitemOpen
  \bibfield  {author} {\bibinfo {author} {\bibfnamefont {Y.~P.}\ \bibnamefont
  {Korneeva}}, \bibinfo {author} {\bibfnamefont {M.~Y.}\ \bibnamefont
  {Mikhailov}}, \bibinfo {author} {\bibfnamefont {Y.~P.}\ \bibnamefont
  {Pershin}}, \bibinfo {author} {\bibfnamefont {N.~N.}\ \bibnamefont {Manova}},
  \bibinfo {author} {\bibfnamefont {A.~V.}\ \bibnamefont {Divochiy}}, \bibinfo
  {author} {\bibfnamefont {Y.~B.}\ \bibnamefont {Vakhtomin}}, \bibinfo {author}
  {\bibfnamefont {A.~A.}\ \bibnamefont {Korneev}}, \bibinfo {author}
  {\bibfnamefont {K.~V.}\ \bibnamefont {Smirnov}}, \bibinfo {author}
  {\bibfnamefont {A.~G.}\ \bibnamefont {Sivakov}}, \bibinfo {author}
  {\bibfnamefont {A.~Y.}\ \bibnamefont {Devizenko}},\ and\ \bibinfo {author}
  {\bibfnamefont {G.~N.}\ \bibnamefont {Goltsman}},\ }\bibfield  {title}
  {\bibinfo {title} {Superconducting single-photon detector made of {MoSi}
  film},\ }\href {https://doi.org/10.1088/0953-2048/27/9/095012} {\bibfield
  {journal} {\bibinfo  {journal} {Supercond. Sci. Technol.}\ }\textbf {\bibinfo
  {volume} {27}},\ \bibinfo {pages} {095012} (\bibinfo {year}
  {2014})}\BibitemShut {NoStop}%
\bibitem [{\citenamefont {Bartolf}(2016)}]{Bar16boo}%
  \BibitemOpen
  \bibfield  {author} {\bibinfo {author} {\bibfnamefont {H.}~\bibnamefont
  {Bartolf}},\ }\href@noop {} {\emph {\bibinfo {title} {Fluctuation mechanisms
  in superconductors}}}\ (\bibinfo  {publisher} {Springer Spektrum, Springer
  Fachmedien Wiesbaden, Germany},\ \bibinfo {year} {2016})\BibitemShut
  {NoStop}%
\bibitem [{\citenamefont {Clem}\ and\ \citenamefont {Kogan}(2012)}]{Cle12prb}%
  \BibitemOpen
  \bibfield  {author} {\bibinfo {author} {\bibfnamefont {J.~R.}\ \bibnamefont
  {Clem}}\ and\ \bibinfo {author} {\bibfnamefont {V.~G.}\ \bibnamefont
  {Kogan}},\ }\bibfield  {title} {\bibinfo {title} {Kinetic impedance and
  depairing in thin and narrow superconducting films},\ }\href
  {https://doi.org/10.1103/PhysRevB.86.174521} {\bibfield  {journal} {\bibinfo
  {journal} {Phys. Rev. B}\ }\textbf {\bibinfo {volume} {86}},\ \bibinfo
  {pages} {174521} (\bibinfo {year} {2012})}\BibitemShut {NoStop}%
\bibitem [{\citenamefont {Romijn}\ \emph {et~al.}(1982)\citenamefont {Romijn},
  \citenamefont {Klapwijk}, \citenamefont {Renne},\ and\ \citenamefont
  {Mooij}}]{Rom82prb}%
  \BibitemOpen
  \bibfield  {author} {\bibinfo {author} {\bibfnamefont {J.}~\bibnamefont
  {Romijn}}, \bibinfo {author} {\bibfnamefont {T.~M.}\ \bibnamefont
  {Klapwijk}}, \bibinfo {author} {\bibfnamefont {M.~J.}\ \bibnamefont
  {Renne}},\ and\ \bibinfo {author} {\bibfnamefont {J.~E.}\ \bibnamefont
  {Mooij}},\ }\bibfield  {title} {\bibinfo {title} {Critical pair-breaking
  current in superconducting aluminum strips far below ${T}_{c}$},\ }\href
  {https://doi.org/10.1103/PhysRevB.26.3648} {\bibfield  {journal} {\bibinfo
  {journal} {Phys. Rev. B}\ }\textbf {\bibinfo {volume} {26}},\ \bibinfo
  {pages} {3648} (\bibinfo {year} {1982})}\BibitemShut {NoStop}%
\bibitem [{\citenamefont {Zhang}\ \emph {et~al.}(2019)\citenamefont {Zhang},
  \citenamefont {Lita}, \citenamefont {Smirnov}, \citenamefont {Liu},
  \citenamefont {Zhu}, \citenamefont {Verma}, \citenamefont {Nam},\ and\
  \citenamefont {Schilling}}]{Zha19arx}%
  \BibitemOpen
  \bibfield  {author} {\bibinfo {author} {\bibfnamefont {X.}~\bibnamefont
  {Zhang}}, \bibinfo {author} {\bibfnamefont {A.~E.}\ \bibnamefont {Lita}},
  \bibinfo {author} {\bibfnamefont {K.}~\bibnamefont {Smirnov}}, \bibinfo
  {author} {\bibfnamefont {H.}~\bibnamefont {Liu}}, \bibinfo {author}
  {\bibfnamefont {D.}~\bibnamefont {Zhu}}, \bibinfo {author} {\bibfnamefont
  {V.~B.}\ \bibnamefont {Verma}}, \bibinfo {author} {\bibfnamefont {S.~W.}\
  \bibnamefont {Nam}},\ and\ \bibinfo {author} {\bibfnamefont {A.}~\bibnamefont
  {Schilling}},\ }\bibfield  {title} {\bibinfo {title} {Strong suppression of
  the resistivity near the transition to superconductivity in narrow
  micro-bridges in external magnetic fields},\ }\href@noop {} {\bibfield
  {journal} {\bibinfo  {journal} {e-print}\ ,\ \bibinfo {eid}
  {arXiv:1909.02915}} (\bibinfo {year} {2019})},\ \Eprint
  {https://arxiv.org/abs/1909.02915} {arXiv:1909.02915 [cond-mat.supr-con]}
  \BibitemShut {NoStop}%
\bibitem [{\citenamefont {Ceccarelli}\ \emph {et~al.}(2019)\citenamefont
  {Ceccarelli}, \citenamefont {Vasyukov}, \citenamefont {Wyss}, \citenamefont
  {Romagnoli}, \citenamefont {Rossi}, \citenamefont {Moser},\ and\
  \citenamefont {Poggio}}]{Cec19prb}%
  \BibitemOpen
  \bibfield  {author} {\bibinfo {author} {\bibfnamefont {L.}~\bibnamefont
  {Ceccarelli}}, \bibinfo {author} {\bibfnamefont {D.}~\bibnamefont
  {Vasyukov}}, \bibinfo {author} {\bibfnamefont {M.}~\bibnamefont {Wyss}},
  \bibinfo {author} {\bibfnamefont {G.}~\bibnamefont {Romagnoli}}, \bibinfo
  {author} {\bibfnamefont {N.}~\bibnamefont {Rossi}}, \bibinfo {author}
  {\bibfnamefont {L.}~\bibnamefont {Moser}},\ and\ \bibinfo {author}
  {\bibfnamefont {M.}~\bibnamefont {Poggio}},\ }\bibfield  {title} {\bibinfo
  {title} {Imaging pinning and expulsion of individual superconducting vortices
  in amorphous {MoSi} thin films},\ }\href
  {https://doi.org/10.1103/PhysRevB.100.104504} {\bibfield  {journal} {\bibinfo
   {journal} {Phys. Rev. B}\ }\textbf {\bibinfo {volume} {100}},\ \bibinfo
  {pages} {104504} (\bibinfo {year} {2019})}\BibitemShut {NoStop}%
\bibitem [{\citenamefont {Lusche}\ \emph {et~al.}(2014)\citenamefont {Lusche},
  \citenamefont {Semenov}, \citenamefont {Korneeva}, \citenamefont {Trifonov},
  \citenamefont {Korneev}, \citenamefont {Gol'tsman},\ and\ \citenamefont
  {H\"ubers}}]{Lus14prb}%
  \BibitemOpen
  \bibfield  {author} {\bibinfo {author} {\bibfnamefont {R.}~\bibnamefont
  {Lusche}}, \bibinfo {author} {\bibfnamefont {A.}~\bibnamefont {Semenov}},
  \bibinfo {author} {\bibfnamefont {Y.}~\bibnamefont {Korneeva}}, \bibinfo
  {author} {\bibfnamefont {A.}~\bibnamefont {Trifonov}}, \bibinfo {author}
  {\bibfnamefont {A.}~\bibnamefont {Korneev}}, \bibinfo {author} {\bibfnamefont
  {G.}~\bibnamefont {Gol'tsman}},\ and\ \bibinfo {author} {\bibfnamefont
  {H.-W.}\ \bibnamefont {H\"ubers}},\ }\bibfield  {title} {\bibinfo {title}
  {Effect of magnetic field on the photon detection in thin superconducting
  meander structures},\ }\href {https://doi.org/10.1103/PhysRevB.89.104513}
  {\bibfield  {journal} {\bibinfo  {journal} {Phys. Rev. B}\ }\textbf {\bibinfo
  {volume} {89}},\ \bibinfo {pages} {104513} (\bibinfo {year}
  {2014})}\BibitemShut {NoStop}%
\bibitem [{\citenamefont {Engel}\ \emph {et~al.}(2012)\citenamefont {Engel},
  \citenamefont {Schilling}, \citenamefont {Il'in},\ and\ \citenamefont
  {Siegel}}]{Eng12prb}%
  \BibitemOpen
  \bibfield  {author} {\bibinfo {author} {\bibfnamefont {A.}~\bibnamefont
  {Engel}}, \bibinfo {author} {\bibfnamefont {A.}~\bibnamefont {Schilling}},
  \bibinfo {author} {\bibfnamefont {K.}~\bibnamefont {Il'in}},\ and\ \bibinfo
  {author} {\bibfnamefont {M.}~\bibnamefont {Siegel}},\ }\bibfield  {title}
  {\bibinfo {title} {Dependence of count rate on magnetic field in
  superconducting thin-film {TaN} single-photon detectors},\ }\href
  {https://doi.org/10.1103/PhysRevB.86.140506} {\bibfield  {journal} {\bibinfo
  {journal} {Phys. Rev. B}\ }\textbf {\bibinfo {volume} {86}},\ \bibinfo
  {pages} {140506} (\bibinfo {year} {2012})}\BibitemShut {NoStop}%
\bibitem [{\citenamefont {Ilin}\ \emph {et~al.}(2014)\citenamefont {Ilin},
  \citenamefont {Henrich}, \citenamefont {Luck}, \citenamefont {Liang},
  \citenamefont {Siegel},\ and\ \citenamefont {Vodolazov}}]{Ili14prb}%
  \BibitemOpen
  \bibfield  {author} {\bibinfo {author} {\bibfnamefont {K.}~\bibnamefont
  {Ilin}}, \bibinfo {author} {\bibfnamefont {D.}~\bibnamefont {Henrich}},
  \bibinfo {author} {\bibfnamefont {Y.}~\bibnamefont {Luck}}, \bibinfo {author}
  {\bibfnamefont {Y.}~\bibnamefont {Liang}}, \bibinfo {author} {\bibfnamefont
  {M.}~\bibnamefont {Siegel}},\ and\ \bibinfo {author} {\bibfnamefont {D.~Y.}\
  \bibnamefont {Vodolazov}},\ }\bibfield  {title} {\bibinfo {title} {Critical
  current of {Nb}, {NbN}, and {TaN} thin-film bridges with and without
  geometrical nonuniformities in a magnetic field},\ }\href
  {https://doi.org/10.1103/PhysRevB.89.184511} {\bibfield  {journal} {\bibinfo
  {journal} {Phys. Rev. B}\ }\textbf {\bibinfo {volume} {89}},\ \bibinfo
  {pages} {184511} (\bibinfo {year} {2014})}\BibitemShut {NoStop}%
\bibitem [{\citenamefont {Plourde}\ \emph {et~al.}(2001)\citenamefont
  {Plourde}, \citenamefont {Van~Harlingen}, \citenamefont {Vodolazov},
  \citenamefont {Besseling}, \citenamefont {Hesselberth},\ and\ \citenamefont
  {Kes}}]{Plo01prb}%
  \BibitemOpen
  \bibfield  {author} {\bibinfo {author} {\bibfnamefont {B.~L.~T.}\
  \bibnamefont {Plourde}}, \bibinfo {author} {\bibfnamefont {D.~J.}\
  \bibnamefont {Van~Harlingen}}, \bibinfo {author} {\bibfnamefont {D.~Y.}\
  \bibnamefont {Vodolazov}}, \bibinfo {author} {\bibfnamefont {R.}~\bibnamefont
  {Besseling}}, \bibinfo {author} {\bibfnamefont {M.~B.~S.}\ \bibnamefont
  {Hesselberth}},\ and\ \bibinfo {author} {\bibfnamefont {P.~H.}\ \bibnamefont
  {Kes}},\ }\bibfield  {title} {\bibinfo {title} {Influence of edge barriers on
  vortex dynamics in thin weak-pinning superconducting strips},\ }\href
  {https://doi.org/10.1103/PhysRevB.64.014503} {\bibfield  {journal} {\bibinfo
  {journal} {Phys. Rev. B}\ }\textbf {\bibinfo {volume} {64}},\ \bibinfo
  {pages} {014503} (\bibinfo {year} {2001})}\BibitemShut {NoStop}%
\bibitem [{\citenamefont {Kupriyanov}\ and\ \citenamefont
  {Likharev}(1975)}]{Kup75pss}%
  \BibitemOpen
  \bibfield  {author} {\bibinfo {author} {\bibfnamefont {M.~Y.}\ \bibnamefont
  {Kupriyanov}}\ and\ \bibinfo {author} {\bibfnamefont {K.~K.}\ \bibnamefont
  {Likharev}},\ }\bibfield  {title} {\bibinfo {title} {Effect of an edge
  barrier on the critical current of a superconducting film},\ }\href@noop {}
  {\bibfield  {journal} {\bibinfo  {journal} {Sov. Phys. Solid State}\ }\textbf
  {\bibinfo {volume} {16}},\ \bibinfo {pages} {1835} (\bibinfo {year}
  {1975})}\BibitemShut {NoStop}%
\bibitem [{\citenamefont {Benkraouda}\ and\ \citenamefont
  {Clem}(1998)}]{Ben98prb}%
  \BibitemOpen
  \bibfield  {author} {\bibinfo {author} {\bibfnamefont {M.}~\bibnamefont
  {Benkraouda}}\ and\ \bibinfo {author} {\bibfnamefont {J.~R.}\ \bibnamefont
  {Clem}},\ }\bibfield  {title} {\bibinfo {title} {Critical current from
  surface barriers in {type-II} superconducting strips},\ }\href
  {https://doi.org/10.1103/PhysRevB.58.15103} {\bibfield  {journal} {\bibinfo
  {journal} {Phys. Rev. B}\ }\textbf {\bibinfo {volume} {58}},\ \bibinfo
  {pages} {15103} (\bibinfo {year} {1998})}\BibitemShut {NoStop}%
\bibitem [{\citenamefont {Elistratov}\ \emph {et~al.}(2002)\citenamefont
  {Elistratov}, \citenamefont {Vodolazov}, \citenamefont {Maksimov},\ and\
  \citenamefont {Clem}}]{Eli02prb}%
  \BibitemOpen
  \bibfield  {author} {\bibinfo {author} {\bibfnamefont {A.~A.}\ \bibnamefont
  {Elistratov}}, \bibinfo {author} {\bibfnamefont {D.~Y.}\ \bibnamefont
  {Vodolazov}}, \bibinfo {author} {\bibfnamefont {I.~L.}\ \bibnamefont
  {Maksimov}},\ and\ \bibinfo {author} {\bibfnamefont {J.~R.}\ \bibnamefont
  {Clem}},\ }\bibfield  {title} {\bibinfo {title} {Field-dependent critical
  current in {type-II} superconducting strips: Combined effect of bulk pinning
  and geometrical edge barrier},\ }\href
  {https://doi.org/10.1103/PhysRevB.66.220506} {\bibfield  {journal} {\bibinfo
  {journal} {Phys. Rev. B}\ }\textbf {\bibinfo {volume} {66}},\ \bibinfo
  {pages} {220506} (\bibinfo {year} {2002})}\BibitemShut {NoStop}%
\bibitem [{\citenamefont {Maksimova}\ \emph {et~al.}(2001)\citenamefont
  {Maksimova}, \citenamefont {Zhelezina},\ and\ \citenamefont
  {Maksimov}}]{Mak01epl}%
  \BibitemOpen
  \bibfield  {author} {\bibinfo {author} {\bibfnamefont {G.~M.}\ \bibnamefont
  {Maksimova}}, \bibinfo {author} {\bibfnamefont {N.~V.}\ \bibnamefont
  {Zhelezina}},\ and\ \bibinfo {author} {\bibfnamefont {I.~L.}\ \bibnamefont
  {Maksimov}},\ }\bibfield  {title} {\bibinfo {title} {Critical current and
  negative magnetoresistance of superconducting film with edge barrier},\
  }\href {https://doi.org/10.1209/epl/i2001-00200-6} {\bibfield  {journal}
  {\bibinfo  {journal} {Europhys. Lett.}\ }\textbf {\bibinfo {volume} {53}},\
  \bibinfo {pages} {639} (\bibinfo {year} {2001})}\BibitemShut {NoStop}%
\bibitem [{\citenamefont {Andratskii}\ \emph {et~al.}(1974)\citenamefont
  {Andratskii}, \citenamefont {Grundel'}, \citenamefont {Gubankov},\ and\
  \citenamefont {Pavlov}}]{And74etp}%
  \BibitemOpen
  \bibfield  {author} {\bibinfo {author} {\bibfnamefont {V.~P.}\ \bibnamefont
  {Andratskii}}, \bibinfo {author} {\bibfnamefont {L.~M.}\ \bibnamefont
  {Grundel'}}, \bibinfo {author} {\bibfnamefont {V.~N.}\ \bibnamefont
  {Gubankov}},\ and\ \bibinfo {author} {\bibfnamefont {N.~B.}\ \bibnamefont
  {Pavlov}},\ }\bibfield  {title} {\bibinfo {title} {Destruction of
  superconductivity in thin narrow films by a current},\ }\href@noop {}
  {\bibfield  {journal} {\bibinfo  {journal} {Sov. Phys. JETP}\ }\textbf
  {\bibinfo {volume} {38}},\ \bibinfo {pages} {794} (\bibinfo {year}
  {1974})}\BibitemShut {NoStop}%
\bibitem [{\citenamefont {Vodolazov}\ \emph
  {et~al.}(2015{\natexlab{b}})\citenamefont {Vodolazov}, \citenamefont {Ilin},
  \citenamefont {Merker},\ and\ \citenamefont {Siegel}}]{Vod15sst}%
  \BibitemOpen
  \bibfield  {author} {\bibinfo {author} {\bibfnamefont {D.~Y.}\ \bibnamefont
  {Vodolazov}}, \bibinfo {author} {\bibfnamefont {K.}~\bibnamefont {Ilin}},
  \bibinfo {author} {\bibfnamefont {M.}~\bibnamefont {Merker}},\ and\ \bibinfo
  {author} {\bibfnamefont {M.}~\bibnamefont {Siegel}},\ }\bibfield  {title}
  {\bibinfo {title} {Defect-controlled vortex generation in current-carrying
  narrow superconducting strips},\ }\href
  {https://doi.org/10.1088/0953-2048/29/2/025002} {\bibfield  {journal}
  {\bibinfo  {journal} {Supercond. Sci. Technol.}\ }\textbf {\bibinfo {volume}
  {29}},\ \bibinfo {pages} {025002} (\bibinfo {year}
  {2015}{\natexlab{b}})}\BibitemShut {NoStop}%
\bibitem [{\citenamefont {Samoilov}\ \emph {et~al.}(1995)\citenamefont
  {Samoilov}, \citenamefont {Konczykowski}, \citenamefont {Yeh}, \citenamefont
  {Berry},\ and\ \citenamefont {Tsuei}}]{Sam95prl}%
  \BibitemOpen
  \bibfield  {author} {\bibinfo {author} {\bibfnamefont {A.~V.}\ \bibnamefont
  {Samoilov}}, \bibinfo {author} {\bibfnamefont {M.}~\bibnamefont
  {Konczykowski}}, \bibinfo {author} {\bibfnamefont {N.~C.}\ \bibnamefont
  {Yeh}}, \bibinfo {author} {\bibfnamefont {S.}~\bibnamefont {Berry}},\ and\
  \bibinfo {author} {\bibfnamefont {C.~C.}\ \bibnamefont {Tsuei}},\ }\bibfield
  {title} {\bibinfo {title} {Electric-field-induced electronic instability in
  amorphous {${\mathrm{Mo}}_{3}$Si} superconducting films},\ }\href
  {https://doi.org/10.1103/PhysRevLett.75.4118} {\bibfield  {journal} {\bibinfo
   {journal} {Phys. Rev. Lett.}\ }\textbf {\bibinfo {volume} {75}},\ \bibinfo
  {pages} {4118} (\bibinfo {year} {1995})}\BibitemShut {NoStop}%
\bibitem [{\citenamefont {Bartolf}\ \emph {et~al.}(2010)\citenamefont
  {Bartolf}, \citenamefont {Engel}, \citenamefont {Schilling}, \citenamefont
  {Il'in}, \citenamefont {Siegel}, \citenamefont {H\"ubers},\ and\
  \citenamefont {Semenov}}]{Bar10prb}%
  \BibitemOpen
  \bibfield  {author} {\bibinfo {author} {\bibfnamefont {H.}~\bibnamefont
  {Bartolf}}, \bibinfo {author} {\bibfnamefont {A.}~\bibnamefont {Engel}},
  \bibinfo {author} {\bibfnamefont {A.}~\bibnamefont {Schilling}}, \bibinfo
  {author} {\bibfnamefont {K.}~\bibnamefont {Il'in}}, \bibinfo {author}
  {\bibfnamefont {M.}~\bibnamefont {Siegel}}, \bibinfo {author} {\bibfnamefont
  {H.-W.}\ \bibnamefont {H\"ubers}},\ and\ \bibinfo {author} {\bibfnamefont
  {A.}~\bibnamefont {Semenov}},\ }\bibfield  {title} {\bibinfo {title}
  {Current-assisted thermally activated flux liberation in ultrathin
  nanopatterned {NbN} superconducting meander structures},\ }\href
  {https://doi.org/10.1103/PhysRevB.81.024502} {\bibfield  {journal} {\bibinfo
  {journal} {Phys. Rev. B}\ }\textbf {\bibinfo {volume} {81}},\ \bibinfo
  {pages} {024502} (\bibinfo {year} {2010})}\BibitemShut {NoStop}%
\bibitem [{\citenamefont {Zotova}\ and\ \citenamefont
  {Vodolazov}(2014)}]{Zot14sst}%
  \BibitemOpen
  \bibfield  {author} {\bibinfo {author} {\bibfnamefont {A.~N.}\ \bibnamefont
  {Zotova}}\ and\ \bibinfo {author} {\bibfnamefont {D.~Y.}\ \bibnamefont
  {Vodolazov}},\ }\bibfield  {title} {\bibinfo {title} {Intrinsic detection
  efficiency of superconducting nanowire single photon detector in the modified
  hot spot model},\ }\href {https://doi.org/10.1088/0953-2048/27/12/125001}
  {\bibfield  {journal} {\bibinfo  {journal} {Supercond. Sci. Technol.}\
  }\textbf {\bibinfo {volume} {27}},\ \bibinfo {pages} {125001} (\bibinfo
  {year} {2014})}\BibitemShut {NoStop}%
\bibitem [{\citenamefont {Engel}\ \emph {et~al.}(2015)\citenamefont {Engel},
  \citenamefont {Lonsky}, \citenamefont {Zhang},\ and\ \citenamefont
  {Schilling}}]{Eng15tas}%
  \BibitemOpen
  \bibfield  {author} {\bibinfo {author} {\bibfnamefont {A.}~\bibnamefont
  {Engel}}, \bibinfo {author} {\bibfnamefont {J.}~\bibnamefont {Lonsky}},
  \bibinfo {author} {\bibfnamefont {X.}~\bibnamefont {Zhang}},\ and\ \bibinfo
  {author} {\bibfnamefont {A.}~\bibnamefont {Schilling}},\ }\bibfield  {title}
  {\bibinfo {title} {Detection mechanism in {SNSPD}: Numerical results of a
  conceptually simple, yet powerful detection model},\ }\href
  {https://doi.org/10.1109/TASC.2014.2371537} {\bibfield  {journal} {\bibinfo
  {journal} {IEEE Trans. Appl. Supercond.}\ }\textbf {\bibinfo {volume} {25}},\
  \bibinfo {pages} {2200407} (\bibinfo {year} {2015})}\BibitemShut {NoStop}%
\bibitem [{\citenamefont {Renema}\ \emph {et~al.}(2015)\citenamefont {Renema},
  \citenamefont {Wang}, \citenamefont {Gaudio}, \citenamefont {Komen},
  \citenamefont {op~'t Hoog}, \citenamefont {Sahin}, \citenamefont {Schilling},
  \citenamefont {van Exter}, \citenamefont {Fiore}, \citenamefont {Engel},\
  and\ \citenamefont {de~Dood}}]{Ren15nal}%
  \BibitemOpen
  \bibfield  {author} {\bibinfo {author} {\bibfnamefont {J.~J.}\ \bibnamefont
  {Renema}}, \bibinfo {author} {\bibfnamefont {Q.}~\bibnamefont {Wang}},
  \bibinfo {author} {\bibfnamefont {R.}~\bibnamefont {Gaudio}}, \bibinfo
  {author} {\bibfnamefont {I.}~\bibnamefont {Komen}}, \bibinfo {author}
  {\bibfnamefont {K.}~\bibnamefont {op~'t Hoog}}, \bibinfo {author}
  {\bibfnamefont {D.}~\bibnamefont {Sahin}}, \bibinfo {author} {\bibfnamefont
  {A.}~\bibnamefont {Schilling}}, \bibinfo {author} {\bibfnamefont {M.~P.}\
  \bibnamefont {van Exter}}, \bibinfo {author} {\bibfnamefont {A.}~\bibnamefont
  {Fiore}}, \bibinfo {author} {\bibfnamefont {A.}~\bibnamefont {Engel}},\ and\
  \bibinfo {author} {\bibfnamefont {M.~J.~A.}\ \bibnamefont {de~Dood}},\
  }\bibfield  {title} {\bibinfo {title} {Position-dependent local detection
  efficiency in a nanowire superconducting single-photon detector},\ }\href
  {https://doi.org/10.1021/acs.nanolett.5b01103} {\bibfield  {journal}
  {\bibinfo  {journal} {Nano Lett.}\ }\textbf {\bibinfo {volume} {15}},\
  \bibinfo {pages} {4541} (\bibinfo {year} {2015})}\BibitemShut {NoStop}%
\bibitem [{\citenamefont {Kozorezov}\ \emph {et~al.}(2017)\citenamefont
  {Kozorezov}, \citenamefont {Lambert}, \citenamefont {Marsili}, \citenamefont
  {Stevens}, \citenamefont {Verma}, \citenamefont {Allmaras}, \citenamefont
  {Shaw}, \citenamefont {Mirin},\ and\ \citenamefont {Nam}}]{Koz17prb}%
  \BibitemOpen
  \bibfield  {author} {\bibinfo {author} {\bibfnamefont {A.~G.}\ \bibnamefont
  {Kozorezov}}, \bibinfo {author} {\bibfnamefont {C.}~\bibnamefont {Lambert}},
  \bibinfo {author} {\bibfnamefont {F.}~\bibnamefont {Marsili}}, \bibinfo
  {author} {\bibfnamefont {M.~J.}\ \bibnamefont {Stevens}}, \bibinfo {author}
  {\bibfnamefont {V.~B.}\ \bibnamefont {Verma}}, \bibinfo {author}
  {\bibfnamefont {J.~P.}\ \bibnamefont {Allmaras}}, \bibinfo {author}
  {\bibfnamefont {M.~D.}\ \bibnamefont {Shaw}}, \bibinfo {author}
  {\bibfnamefont {R.~P.}\ \bibnamefont {Mirin}},\ and\ \bibinfo {author}
  {\bibfnamefont {S.~W.}\ \bibnamefont {Nam}},\ }\bibfield  {title} {\bibinfo
  {title} {Fano fluctuations in superconducting-nanowire single-photon
  detectors},\ }\href {https://doi.org/10.1103/PhysRevB.96.054507} {\bibfield
  {journal} {\bibinfo  {journal} {Phys. Rev. B}\ }\textbf {\bibinfo {volume}
  {96}},\ \bibinfo {pages} {054507} (\bibinfo {year} {2017})}\BibitemShut
  {NoStop}%
\bibitem [{\citenamefont {Korneev}\ \emph {et~al.}(2015)\citenamefont
  {Korneev}, \citenamefont {Korneeva}, \citenamefont {Mikhailov}, \citenamefont
  {Pershin}, \citenamefont {Semenov}, \citenamefont {Vodolazov}, \citenamefont
  {Divochiy}, \citenamefont {Vakhtomin}, \citenamefont {Smirnov}, \citenamefont
  {Sivakov}, \citenamefont {Devizenko},\ and\ \citenamefont
  {Goltsman}}]{Kor15tas}%
  \BibitemOpen
  \bibfield  {author} {\bibinfo {author} {\bibfnamefont {A.~A.}\ \bibnamefont
  {Korneev}}, \bibinfo {author} {\bibfnamefont {Y.~P.}\ \bibnamefont
  {Korneeva}}, \bibinfo {author} {\bibfnamefont {M.~Y.}\ \bibnamefont
  {Mikhailov}}, \bibinfo {author} {\bibfnamefont {Y.~P.}\ \bibnamefont
  {Pershin}}, \bibinfo {author} {\bibfnamefont {A.~V.}\ \bibnamefont
  {Semenov}}, \bibinfo {author} {\bibfnamefont {D.~Y.}\ \bibnamefont
  {Vodolazov}}, \bibinfo {author} {\bibfnamefont {A.~V.}\ \bibnamefont
  {Divochiy}}, \bibinfo {author} {\bibfnamefont {Y.~B.}\ \bibnamefont
  {Vakhtomin}}, \bibinfo {author} {\bibfnamefont {K.~V.}\ \bibnamefont
  {Smirnov}}, \bibinfo {author} {\bibfnamefont {A.~G.}\ \bibnamefont
  {Sivakov}}, \bibinfo {author} {\bibfnamefont {A.~Y.}\ \bibnamefont
  {Devizenko}},\ and\ \bibinfo {author} {\bibfnamefont {G.~N.}\ \bibnamefont
  {Goltsman}},\ }\bibfield  {title} {\bibinfo {title} {Characterization of
  {MoSi} superconducting single-photon detectors in the magnetic field},\
  }\href {https://doi.org/10.1109/TASC.2014.2376892} {\bibfield  {journal}
  {\bibinfo  {journal} {IEEE Trans. Appl. Supercond.}\ }\textbf {\bibinfo
  {volume} {25}},\ \bibinfo {pages} {2200504} (\bibinfo {year}
  {2015})}\BibitemShut {NoStop}%
\bibitem [{\citenamefont {Bulaevskii}\ \emph {et~al.}(2012)\citenamefont
  {Bulaevskii}, \citenamefont {Graf},\ and\ \citenamefont {Kogan}}]{Bul12prb}%
  \BibitemOpen
  \bibfield  {author} {\bibinfo {author} {\bibfnamefont {L.~N.}\ \bibnamefont
  {Bulaevskii}}, \bibinfo {author} {\bibfnamefont {M.~J.}\ \bibnamefont
  {Graf}},\ and\ \bibinfo {author} {\bibfnamefont {V.~G.}\ \bibnamefont
  {Kogan}},\ }\bibfield  {title} {\bibinfo {title} {Vortex-assisted photon
  counts and their magnetic field dependence in single-photon superconducting
  detectors},\ }\href {https://doi.org/10.1103/PhysRevB.85.014505} {\bibfield
  {journal} {\bibinfo  {journal} {Phys. Rev. B}\ }\textbf {\bibinfo {volume}
  {85}},\ \bibinfo {pages} {014505} (\bibinfo {year} {2012})}\BibitemShut
  {NoStop}%
\end{thebibliography}
